\documentclass[12pt,epsfig]{article}   
\usepackage{feynmp}
\newcommand{\be}{\begin{equation}}   
\newcommand{\ee}{\end{equation}}   
\newcommand{\beqn}{\begin{eqnarray}}   
\newcommand{\eeqn}{\end{eqnarray}}
\newcommand{\mbf}[1]{\mbox{\boldmath $#1$}} 
\begin{document}
DESY 99-195                           \hfill ISSN 0418--9833 \\
\begin{center}   
\begin{Large}   
\vspace*{0.5cm}  
{\bf A New Odderon Solution in Perturbative QCD}\\   
\end{Large}   
\vspace{0.5cm}   
J. Bartels$^a$ \footnote   
        {Supported by the TMR Network "QCD and Deep Structure of Elementary 
     Particles"},  
L.N.Lipatov$^b$ \footnote{Supported by the CRDF and INTAS-RFBR
(grants:RP1-253,95-0311) and by the Deutsche Forschungsgemeinschaft},
G.P.Vacca$^a$ \footnote{Supported by the Alexander von Humboldt Stiftung}\\
$^a$ II. Inst. f. Theoretische Physik,    
Univ. Hamburg, Luruper Chaussee 149, D-22761 Hamburg\\  
$^b$ St.Petersburg Nuclear Physics Institute,
     Gatchina, Orlova Roscha, 188350, Russia.
\end{center}   
\vspace*{0.75cm} 
%%%%%%%%%%%%%%%%%%%%%%%%%%%%%%%%%%%%%%%%%%%%%%%%%%%%%%%%%%%%%%%%%%%
\begin{abstract}
\noindent
We present and discuss a new bound state solution of the three gluon system
in perturbative QCD. It carries the quantum numbers of the odderon, has 
intercept at one and couples to the impact factor
$\gamma^* \to \eta_c$.
\end{abstract}
%%%%%%%%%%%%%%%%%%%%%%%%%%%%%%%%%%%%%%%%%%%%%%%%%%%%%%%%%%%%%%%%%%%
1. The unitarization of the BFKL Pomeron \cite{BFKL} presents one of the major 
tasks in QCD. After the successful calculation of the NLO corrections to the
BFKL kernel \cite{FL} and recent progress in analyzing its properties 
\cite{NLOint}, there are several directions in going beyond the 
two-gluon ladder approximation. One of them investigates 
multigluon compound states. After the first formulation of the BKP 
equations \cite{BKP} it was found that, in the large-$N_c$ limit, 
their solutions have remarkable mathematical properties ~\cite{Lintegr},
and the hamiltonian is the same as for the integrable spin chain 
\cite{LFKchain}.
The existence of integrals of motion 
~\cite{Lintegr} and the duality symmetry ~\cite{Ldual} provide powerful tools
in analysing the spectrum of energy eigenvalues. Another line of research
investigates the transition between states with different numbers of gluons
~\cite{B,BW,BL,BE}.\\ \\
As the first step beyond the two-gluon system 
(i.e. the BFKL equation) the spectrum
of the three gluon system (odderon \cite{NICOL}) has attracted much
attention recently. Apart from the theoretical interest in understanding the 
dynamics of the n-gluon states with $n>2$, there is the long-standing 
odderon problem which provides interest from the phenomenological side.
After several variational studies an eigenfunction of the integral of motion
~\cite{Lintegr} with the odderon intercept slightly below one was constructed
by Janik and Wosiek \cite{JW} (see also ~\cite{Ldual}) and verified by
Braun et al \cite{BGN}.
From the phenomenological side, a possible signature of the odderon in
deep inelastic scattering at HERA has been investigated by several authors
\cite{Kwie,Odd}. In this context also the coupling of the odderon to
the $\gamma^* \to \eta_c$ vertex has been calculated ~\cite{Kwie}. 
Another piece
of information relevant for the three gluon channel is the Pomeron $\to$ 
two-odderon vertex which has been obtained from an analysis of the
six-gluon state ~\cite{BE}. Its momentum structure coincides with the
momentum dependence found in ~\cite{Kwie}.\\ \\   
In this letter we present a new explicit solution to the three gluon
system which carries the quantum numbers of the odderon and has interecept
one. It is derived from the momentum structure found in ~\cite{BE} and
~\cite{Kwie}. This solution can be interpreted as the
reggeization of a d-reggeon in QCD (the even-signature color octet
reggeon which is degenerate with the odd-signature reggeized gluon), 
which interacts with the reggeized gluon. Our new solution can also be 
obtained by applying a duality transformation to the antisymmetric solution 
found in \cite{LDIS99}. From the phenomenological point of view, this new
solution seems to be more important than the previous one:
its intercept is higher than that of the totally symmetric 
odderon solution of ~\cite{JW}. In the final part of this letter we shall 
explicitely show how our solution couples to the 
$\gamma^* \to \eta_c$-vertex.\\ \\
%%%%%%%%%%%%%%%%%%%%%%%%%%%%%%%%%%%%%%%%%%%%%%%%%%%%%%%%%%%%%%%%%%%%%%%%%%  
2. Let us begin with the coupling of three gluons to external 
particles. In order to be able to apply perturbative QCD we should start
from a virtual photon, $\gamma^*$, which splits into two quarks.
For the elastic impact factor $\gamma^* \to \gamma^*$ it was shown in 
~\cite{BW} that in the t-channel with three gluons the bootstrap property 
of the gluon reduces the number of reggeized gluons to two, 
i.e. there is no state with three reggeized gluons.
Therefore, a nonzero coupling of a three gluon t-channel to external particles
needs an outgoing state
whose parity is even, i.e. opposite to the photon. The easiest candidate 
is the $\gamma^* \to \eta_c$-vertex which has been calculated in 
~\cite{Kwie} (Fig1a). Its momentum structure (as a function of transverse 
momenta) has the following the form:
\beqn
\Phi_{\gamma \to \eta_c}^{(i)} \sim g_s^3 \epsilon_{ij} \frac{q_j}{\mbf{q}^2}
\left( \sum_{(123)} 
\frac{(\mbf{k}_1 + \mbf{k}_2 - \mbf{k}_3) \cdot \mbf{q}}{Q^2+4m_c^2 + 
(\mbf{k}_1 + \mbf{k}_2 - \mbf{k}_3)^2}
-\frac{\mbf{q}^2}{Q^2+4m_c^2 + \mbf{q}^2} \right) \, .
\eeqn
Here $\mbf{q}=\mbf{k}_1+\mbf{k}_2+\mbf{k}_3$, 
the sum extends over the cyclic permutations 
of (1,2,3), and we have left out an overall factor in 
front which is not important for our present discussion. 
The color structure is simply given by the symmetric structure 
constants $d^{a_1 a_2 a_3}$. By introducing the short hand notation
\beqn
\varphi_{-}^{(i)}(\mbf{k},\mbf{k}') = g_s^2
\epsilon_{ij} \frac{q_j}{\mbf{q}^2}
\frac{(\mbf{k}-\mbf{k}')(\mbf{k}+\mbf{k}')}
{Q^2+4 m_c^2 +(\mbf{k}-\mbf{k}')^2} \, ,
\eeqn
we rewrite (1) as
\beqn
\Phi_{\gamma \to \eta_c}^{(i)} \sim g_s \Bigl( \sum_{(123)} 
\varphi_{-}^{(i)}(\mbf{k}_1+\mbf{k}_2, \mbf{k}_3)
- \varphi_{-}^{(i)}(\mbf{k}_1+\mbf{k}_2+\mbf{k}_3,\mbf{0}) \Bigr).
\eeqn
The function $\varphi_{-}$ is antisymmetric under the exchange of its two
arguments. 
It is easy to see that $\Phi$ vanishes as one of the transverse momenta
$\mbf{k}_i$ goes to zero (with fixed $\mbf{q}$).
The full sum of the cyclic permutations 
is symmetric under the exchange of any pair of momenta
$(\mbf{k}_i,\mbf{k}_j)$, but because of the antisymmetry of $\varphi_{-}$ its 
symmetry structure is more involved and will be discussed below.\\ \\
%%%%%%%%%%%%%%%%%%%%%%%%%%%%%%%%%%%%%%%%%%%%%%%%%%%%%%%%%%%%%%%%%%%
%     Figure 1 created with feynmp
%%%%%%%%%%%%%%%%%%%%%%%%%%%%%%%%%%%%%%%%%%%%%%%%%%%%%%%%%%%%%%%%%%%%%%
\begin{fmffile}{fig1}
\begin{figure}
 \begin{center}
  \vspace{1cm}
% First figure
  \begin{fmfgraph*}(40,25)
    \fmfset{arrow_len}{3mm}
    \fmfset{wiggly_len}{3mm}
    \fmfset{curly_len}{2mm}
    \fmfpen{thin}
    \fmfipair{phi,phv,eco,ecv,gav,gbv,gcv,gao,gbo,gco,sigma}
    \fmfiequ{xpart(phi)}{0w}
    \fmfiequ{xpart(eco)}{1w}
    \fmfiequ{ypart(phi)}{0.8h}
    \fmfiequ{xpart(phv)}{0.25w}
    \fmfiequ{xpart(ecv)}{0.75w}
    \fmfiequ{ypart(gao)}{0h}
    \fmfiequ{xpart(sigma)}{-0.1w}
    \fmfiequ{ypart(sigma)}{0.6h}
    \fmfiequ{ypart(phv)}{ypart(phi)}
    \fmfiequ{ypart(phi)}{ypart(eco)}
    \fmfiequ{ypart(ecv)}{ypart(eco)}
    \fmfiequ{ypart(gao)}{ypart(gbo)}
    \fmfiequ{ypart(gao)}{ypart(gco)}
    \fmfiequ{xpart(gao)}{xpart(gav)}
    \fmfiequ{xpart(gbo)}{xpart(gbv)}
    \fmfiequ{xpart(gco)}{xpart(gcv)}

    \fmfi{photon}{phi--phv}
    \fmfi{dashes}{ecv--eco}
    \fmfipath{p[]}
    \fmfiset{p1}{ecv .. tension 2 .. {down}phv}
    \fmfiset{p2}{phv .. tension 2 .. {up}ecv}
    \fmfiequ{gcv}{point length(p1)/4 of p1}
    \fmfiequ{gav}{point length(p2)/4 of p2}
    \fmfiequ{gbv}{point length(p2)/2 of p2}
    \fmfi{plain}{subpath(0,1/4) of p1}
    \fmfi{fermion}{subpath(1/4,1) of p1}
    \fmfi{fermion}{subpath(0,1/4) of p2}
    \fmfi{plain}{subpath(1/4,1/2) of p2}
    \fmfi{plain}{subpath(1/2,1) of p2}
    \fmfi{gluon}{gav--gao}
    \fmfi{gluon}{gbv--gbo}
    \fmfi{gluon,rubout=3}{gcv--gco}
    \fmfiv{d.sh=circle,d.siz=2thick}{ecv}

    \fmfiv{label=$\mbf{\gamma^*}$}{phi}
    \fmfiv{label=$\mbf{\eta_c}$}{eco}
    \fmfiv{label=$\mbf{k_1}$}{gao}
    \fmfiv{label=$\mbf{k_2}$}{gbo}
    \fmfiv{label=$\mbf{k_3}$}{gco}
    \fmfiv{label={\Huge $\Sigma$},lab.ang=-90}{sigma}
    \fmfiv{label=\textbf{(a)},lab.ang=-90}{se}
\end{fmfgraph*}
\hspace{2cm}
  \begin{fmfgraph*}(40,25)
    \fmfset{arrow_len}{3mm}
    \fmfset{wiggly_len}{3mm}
    \fmfset{curly_len}{2mm}
    \fmfpen{thin}
    \fmfipair{gli,glv,gri,grv,gav,gbv,gcv,gao,gbo,gco}
    \fmfipair{gdv,gev,gfv,gdo,geo,gfo,vl,vr}
    \fmfipair{sigma,la,lb,ra,rb}
    \fmfiequ{xpart(gli)}{0.4w}
    \fmfiequ{xpart(gri)}{0.6w}
    \fmfiequ{ypart(gli)}{1h}
    \fmfiequ{ypart(gli)}{ypart(gri)}
    \fmfiequ{xpart(vl)}{0.3w}
    \fmfiequ{ypart(vl)}{0.5h}
    \fmfiequ{xpart(vr)}{0.7w}
    \fmfiequ{ypart(vr)}{ypart(vl)}
    \fmfiequ{ypart(la)}{.05h}
    \fmfiequ{ypart(lb)}{0h}
    \fmfiequ{ypart(la)}{ypart(ra)}
    \fmfiequ{ypart(lb)}{ypart(rb)}
    \fmfiequ{xpart(la)}{.1w}
    \fmfiequ{xpart(lb)}{.3w}
    \fmfiequ{xpart(ra)}{.9w}
    \fmfiequ{xpart(rb)}{.7w}
    \fmfiset{p1}{vr .. tension 2 .. {down}vl}
    \fmfiset{p2}{vl .. tension 2 .. {up}vr}
    \fmfiset{p3}{la .. lb}
    \fmfiset{p4}{ra .. rb}
    \fmfiequ{grv}{point 0.33 length(p1) of p1}
    \fmfiequ{glv}{point 0.66 length(p1) of p1}
    \fmfiequ{gav}{point 0.2 length(p2) of p2}
    \fmfiequ{gbv}{point 0.3 length(p2) of p2}
    \fmfiequ{gcv}{point 0.4 length(p2) of p2}
    \fmfiequ{gdv}{point 0.6 length(p2) of p2}
    \fmfiequ{gev}{point 0.7 length(p2) of p2}
    \fmfiequ{gfv}{point 0.8 length(p2) of p2}
    \fmfiequ{gao}{point 0 length(p3) of p3}
    \fmfiequ{gbo}{point .5 length(p3) of p3}
    \fmfiequ{gco}{point  length(p3) of p3}
    \fmfiequ{gfo}{point 0 length(p4) of p4}
    \fmfiequ{geo}{point .5 length(p4) of p4}
    \fmfiequ{gdo}{point  length(p4) of p4}
    \fmfi{gluon}{gli--glv}
    \fmfi{gluon}{gri--grv}
    \fmfi{gluon}{gav--gao}
    \fmfi{gluon}{gbv--gbo}
    \fmfi{gluon}{gcv--gco}
    \fmfi{gluon}{gdv--gdo}
    \fmfi{gluon}{gev--geo}
    \fmfi{gluon}{gfv--gfo}
    \fmfi{dbl_plain}{subpath(0,1) of p1}
    \fmfi{dbl_plain}{subpath(0,1) of p2}
    \fmfiv{label=\textbf{(b)},lab.ang=-90}{se}   
\end{fmfgraph*}
\vspace{1cm}
\caption{(a) The $\Phi_{\gamma \to \eta_c}^{(i)}$ impact factor corresponds
to the sum of the graphs with different gluon configurations.
 (b) The effective Pomeron $\to$ two-Odderon vertex $W$. Each group of three
outcoming gluons is in a totally symmetric color singlet state. 
}
\end{center}
\end{figure}
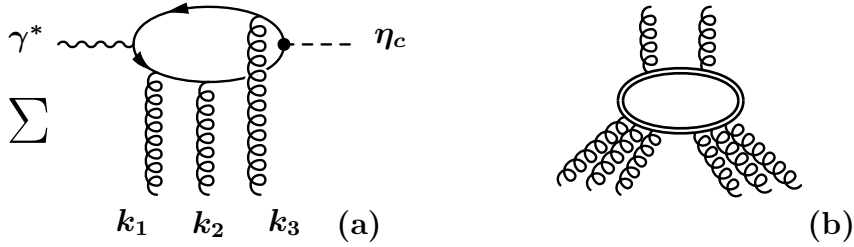

\end{fmffile}
%%%%%%%%%%%%%%%%%%%%%%%%%%%%%%%%%%%%%%%%%%%%%%%%%%%%%%%%%%%%%%%%%%%
%           End of figure
%%%%%%%%%%%%%%%%%%%%%%%%%%%%%%%%%%%%%%%%%%%%%%%%%%%%%%%%%%%%%%%%%%%
Interesting enough, the same momentum structure (3) has also been found
in the new Pomeron $\to$ two-odderon vertex of ~\cite{BE} 
(Fig.1b). Starting from eq.(6.12) in ~\cite{BE}, one first expresses 
~\cite{Ewth} the function 
$W(\mbf{k}_1,\mbf{k}_2,\mbf{k}_3;\mbf{k}_4,\mbf{k}_5,\mbf{k}_6)$ 
in terms of the function 
$G(\mbf{k}_1, \mbf{k}_2, \mbf{k}_3)$ which was first introduced in 
~\cite{BW} and further investigated in
~\cite{Vacca}. As an important property we note that this $G$-function 
vanishes if either $\mbf{k}_1$ or $\mbf{k}_3$ goes to zero. 
Next we introduce the function
\beqn
\varphi_{--}(\mbf{k}_1,\mbf{k}_2;\mbf{k}_3,\mbf{k}_4)= \hspace{5cm}\nonumber \\
g_s^4 \Bigl( G(\mbf{k}_1,\mbf{k}_2+\mbf{k}_3,\mbf{k}_4)-
G(\mbf{k}_2,\mbf{k}_1+\mbf{k}_3,\mbf{k}_4)
-G(\mbf{k}_1,\mbf{k}_2+\mbf{k}_4,\mbf{k}_3) + 
G(\mbf{k}_2,\mbf{k}_1+\mbf{k}_4,\mbf{k}_3) \Bigr).
\eeqn
Then the Pomeron $\to$ two-odderon vertex $W$ takes the following 
form:
\beqn
W(\mbf{k}_1,\mbf{k}_2,\mbf{k}_3;\mbf{k}_4,\mbf{k}_5,\mbf{k}_6) 
\sim \hspace{4cm} \nonumber \\
\sum_{(123)} \sum_{(456)} 
\varphi_{--}(\mbf{k}_1+\mbf{k}_2,\mbf{k}_3;\mbf{k}_4,\mbf{k}_5+\mbf{k}_6)
- \sum_{(123)} 
\varphi_{--} (\mbf{k}_1+\mbf{k}_2,\mbf{k}_3;\mbf{0},\mbf{k}_4+\mbf{k}_5+\mbf{k}_6)
 \nonumber \\
- \sum_{(456)} 
\varphi_{--} (\mbf{k}_1+\mbf{k}_2+\mbf{k}_3,\mbf{0};\mbf{k}_4,\mbf{k}_5+\mbf{k}_6)
+  
\varphi_{--}(\mbf{k}_1+\mbf{k}_2+\mbf{k}_3,\mbf{0};\mbf{0},\mbf{k}_4+\mbf{k}_5+\mbf{k}_6)
\eeqn
Now one easily sees that the momentum structure of the
three gluon systems (123) or (456) is the same as in (3). Again we have the
property that $W$ vanishes as any of the $\mbf{k}_i$ goes to zero 
(from (5) one
sees immediately that $W\to 0$ when 
$\mbf{k}_i\to 0$, with the total odderon momenta 
$\mbf{k}_1+\mbf{k}_2+\mbf{k}_3$ and $\mbf{k}_4+\mbf{k}_5+\mbf{k}_6$ 
being kept fixed). The color structure is given by the product of the 
two d-tensors: $d^{a_1 a_2 a_3} 
d^{a_4 a_5 a_6}$.
\\ \\
%%%%%%%%%%%%%%%%%%%%%%%%%%%%%%%%%%%%%%%%%%%%%%%%%%%%%%%%%%%%%%%%%%%%%%
3. Starting from this momentum structure it is easy to find a solution for
the three gluon system. For simplicity we return to the impact factor
$\Phi_{\gamma \to \eta_c}$ in (3).
Disregarding, for the moment, the last term which
serves as a subtraction constant, we consider the convolution of 
$\sum_{(123)} \varphi_{-}(\mbf{k}_1+\mbf{k}_2, \mbf{k}_3)$
with the kernel for the three gluon state:
\beqn
K_{(123)}= \sum_{(ij)} K_{(ij)}
\eeqn
where
\beqn
K_{(12)}=K(\mbf{k}_1,\mbf{k}_2;\mbf{k}'_1,\mbf{k}'_2)
\eeqn
is the LO BFKL kernel which includes the gluon trajectory functions.
We find
\beqn
\left(K_{(123)} \otimes  \frac{1}{\mbf{k}_1^2\mbf{k}_2^2\mbf{k}_3^2}
\sum_{(123)} \varphi_{-}(\mbf{k}_1+\mbf{k}_2, \mbf{k}_3) \right)
\left(\mbf{k}_1,\mbf{k}_2,\mbf{k}_3\right) 
= \nonumber \\
\sum_{(123)} \left( K_{(12)} \otimes 
\frac{1}{\mbf{k}_1^2\mbf{k}_2^2}
\varphi_{-}({\mbf{k}_1,\mbf{k}_2}) \right)\left( \mbf{k}_1+\mbf{k}_2,\mbf{k}_3 \right)
\eeqn
In deriving this result it is important to use the color structure
$d^{a_1a_2a_3}$, the antisymmetry of $\varphi_{-}$, and the
bootstrap property of the BFKL kernel.
The latter is a relation which guarantees that production amplitudes
with the gluon quantum number in their $t$ channels used for the
construction of the absortive part are characterized by just a single
reggeized gluon exchange (at leading and next-to-leading orders).
The convolution symbol $\otimes$ denotes the integral over
transverse momenta, and we have explicitly written the gluon propagators 
between $\varphi_{-}$ and the BFKL kernel. For the moment we ignore the fact 
that the integral in (8)  
is infrared singular, since the function $\varphi_{-}$
does not vanish as one of its argument goes to zero. Next we replace
the function $\varphi_{-}(\mbf{k},\mbf{q}-\mbf{k})$ by the BFKL (normalized)
eigenfunction $E^{(\nu,n)}(\mbf{k},\mbf{q}-\mbf{k})$;
for odd values of the conformal spin $n$ this function 
is odd under the interchange of its arguments $\mbf{k}$ and
$\mbf{q}-\mbf{k}$. This leads to the following definition:
\beqn
E_3^{(\nu,n)}(\mbf{k}_1,\mbf{k}_2,\mbf{k}_3)=g_s \frac{N_c}{\sqrt{N_c^2-4}}
\frac{1}{\sqrt{-3\chi(\nu,n)}}
\sum_{(123)} \frac{(\mbf{k}_1+\mbf{k}_2)^2}{\mbf{k}_1^2\mbf{k}_2^2} 
E^{(\nu,n)}(\mbf{k}_1+\mbf{k}_2,\mbf{k}_3), 
\eeqn 
where 
\beqn
\chi(\nu,n)=\frac{N_c\alpha_s}{\pi} 
\left( 2\psi(1)-\psi(\frac{1+|n|}{2}+i\nu)
                  -\psi(\frac{1+|n|}{2}-i\nu) \right)
\eeqn
is the characteristic function of the BFKL kernel, and the global color
structure is again given by  $d^{a_1 a_2 a_3}$.
The function $E_3^{(\nu,n)}$ satisfies (8), but since $E^{(\nu,n)}$ 
is an eigenfunction of the BFKL kernel, we can go one step further and obtain 
\beqn
K_{(123)}\otimes E_3^{(\nu,n)} = \chi(\nu,n) E_3^{(\nu,n)}.
\eeqn 
The leading eigenvalue for $n=\pm1$, $\nu=0$ lies at
zero, i.e. in the angular momentum plane the rightmost singularity lies at 
$j=1$. Hence this solution dominates over the totally symmetric solution
of \cite{JW}.
Let us remark that in (9) we have included a normalization
factor such that the norm of $E_3^{(\nu,n)}$ turns out to be
equal to the norm of $E^{(\nu,n)}$.\\ \\
Property (8) can be viewed as the reggeization of the d-reggeon: starting 
from the initial condition (as given by (3)) and evolving the three 
gluon state with the help of the kernel (6), the identity (8) tells us that 
the three gluon system ''collapses'' into
a two-reggeon state, where one reggeon is the well-known reggeized gluon
(in the antisymmetric color octet reperesentation), the other one
a d-reggeon (belonging to the symmetric color octet representation). 
The full state is in a color singlet, but it has odd C-parity.
This situation can be compared with the three gluon state in the Pomeron
channel (even C-parity) discussed in ~\cite{B,BW}: here the initial condition
is given by the $D_{(3,0)}$ function. The three gluons also evolve and 
''collapse'' into two reggeized gluons. The main difference lies in the 
evolution which, in the Pomeron channel, contains also a transition kernel 
$2 \to 3$ gluons. Such a kernel is absent in the odderon channel.\\ \\
4. For our further discussion it is convenient to switch to 
configuration space. We will show that the new solution (9) can be also 
be obtained from another solution which has been 
found recently in  \cite{LDIS99}. Using the Moebius invariance of the 
Hamiltonian for the compound states
$
\Psi _{m,\widetilde{m}}(\mbf{\rho }_{0})$ of the three
reggeized
gluons in LLA in the impact parameter space $\mbf{\rho }$, we
can write the ansatz for the corresponding wave function
\begin{equation}
f_{m,\widetilde{m}}(\mbf{\rho }_{1},\,\mbf{\rho }%
_{2},\,\mbf{\rho }_{3};\,\mbf{\rho }_{0})=\left(
\frac{%
\rho _{23}}{\rho _{20}\rho _{30}}\right) ^{m}\left( \frac{\rho
_{23}^{*}}{%
\rho _{20}^{*}\rho _{30}^{*}}\right) ^{\widetilde{m}}\varphi
_{m,\widetilde{m%
}}(x,x^{*})\,.
\end{equation}
Here $\rho _{kl}=\rho _{k}-\rho _{l}$, $\rho _{k}=\rho _{k}^{1}+i\rho
_{k}^{2}$, and $x=x^{1}+ix^{2}$ is the anharmonic ratio:
 \begin{equation}
x=\frac{\rho _{12}\rho _{30}}{\rho _{10}\rho _{32}}\,.\,
\end{equation}
The quantum numbers $m$ and $\widetilde{m}$ are the conformal
weights of the state $\Psi _{m,\widetilde{m}}(\mbf{\rho }_{0})$
belonging to the basic series of the unitary representations of the 
Moebius group:
\begin{equation}
m=\frac{1}{2}-i\nu +\frac{n}{2}\,,\,\,\widetilde{m}=\frac{1}{2}-i\nu
-\frac{n%
}{2}\,,
\end{equation}
where $n$ is the conformal spin, and $d=1-2i\nu $ is the anomalous dimension
of the operator $O_{m,\widetilde{m}}(\mbf{\rho }_{0})$ describing
the compound state \cite{Lcft}.
The function $f_{m,\widetilde{m}}(\mbf{\rho }_{1},\,%
\mbf{\rho }_{2},\,\mbf{\rho
}_{3};\,\mbf{%
\rho }_{0})$ is an eigenfunction of the integrals of motion $A$ and
$A^{*}$
, where $A=$ $i^{3}\rho _{12}\rho _{23}\rho _{31}\partial _{1}\partial
_{2}\partial _{3}$ with $\partial _{k}=\partial /(\partial \rho _{k})$
\cite{Ldual}.
In the $x$-representation the eigenvalue equation for $A$ takes the form
 
\begin{equation}
A\,_{m}\varphi _{m,\widetilde{m}}(x,x^{*})=\lambda
_{m}\,\,\varphi _{m,\widetilde{m}}(x,x^{*})\,\,,\,
\end{equation}
where $A_{m}$ can be written in the factorized form \cite{Ldual,LDIS99}
 
\begin{equation}
A_{m}=a_{1-m}(x)\,a_{m}(x)\,,\,\,a_{m}(x)=x(1-x)\,(i\partial )^{1+m}\,\,.
\end{equation}
Note, that $A_{m}$ is the ordinary differential operator of the third
order
\begin{equation}
A_{m}=i^{3}x(1-x)\left( x(1-x)\,\partial ^{3}+(2-m)\,(1-2x)\,\partial
^{2}-(2-m)(1-m)\partial \right) \,.
\end{equation}
 
We are looking for a solution which is annihilated by the operator $AA^*$.
The zero modes of the operator $A_m$ with $\lambda _{m}=0$ are $1,\,x^{m}$ and
$(1-x)^{m}$. The corresponding wave function in the $(x,x^{*})$
representation for the state symmetric under the cyclic transmutations 
$\mbf{\rho}_{1}\rightarrow \mbf{\rho}_{2}\rightarrow
\mbf{\rho}_{3}\rightarrow \mbf{\rho}_{1}$ of the
gluon coordinates is
\begin{equation}
\varphi _{m,\widetilde{m}}^{0}(\mbf{x})=1+(-x)^{m}(-x^{*})^{%
\widetilde{m}}+(x-1)^{m}(x^{*}-1)^{\widetilde{m}}.
\end{equation}
For even values of the conformal spin this wave function is not
normalized
and does not correspond to any physical state.
 
However, for odd conformal spins
 
\begin{equation}
n=m-\widetilde{m}=2k+1
\end{equation}
the above expression vanishes at $x \rightarrow 0,1$ and
$%
\infty $
 
\begin{equation}
\varphi _{m,\widetilde{m}}^{0}(x,x^{*})\rightarrow 0\,,
\end{equation}
and is normalized \cite{LDIS99}. In this case the wave function
$f_{m,\widetilde{m}}(%
\mbf{\rho }_{1},\,\mbf{\rho
}_{2},\,\mbf{%
\rho }_{3};\,\mbf{\rho }_{0})$ is anti-symmetric under the
pair
transmutations $\mbf{\rho }_{i}\longleftrightarrow
\mbf{\rho }_{k}$ of the gluon coordinates. Therefore, due to
requirements of the Bose symmetry of the total wave function, it describes a
colourless state with the colour wave function proportional to the
structure
constant $f_{a_{1}a_{2}a_{3}}$ of the gauge group. This state has 
positive charge parity and gives a non-vanishing contribution to the
structure function $g_{2}$ for the deep-inelastic scattering of the
polarized electron off the polarized proton \cite{LDIS99}.
The value of the energy
turns out to be a half of the energy for the Pomeron with corresponding
$m$
and $\widetilde{m}$\thinspace . The minimal value of it is reached for $%
m=1,\,\widetilde{m}=0$ or $m=0,\,\widetilde{m}=1$ and equals zero
\cite{LDIS99}.
\\ \\ 
For the interactions of three (and more) gluons the hamiltonian and the 
integrals of motion are invariant under duality transformations among the
coordinates and momenta of the gluons \cite{Ldual}. Generally this invariance 
leads to a
degeneracy of the spectrum of these operators. Namelly, two 
eigenfunctions $%
\varphi _{m,\widetilde{m}}^{1}(x,x^{*})$ and $\varphi
_{1-m,1-%
\widetilde{m}}^{2}(x,x^{*})$ corresponding to the same
eigenvalues of $A$ and $A^{*}$ are related by the duality operator
$Q_{m,\widetilde{m}}$ \cite{Ldual}:
 
\[
Q_{m,\widetilde{m}}\varphi _{m,\widetilde{m}}^{1}(x,x^{*}%
)=a_{m}(x)\,a_{\widetilde{m}}(x^{*})\,\varphi _{m,\widetilde{m}}^{1}(%
x,x^{*})=
\]
\begin{equation}
\left| x(1-x)\right| ^{2}(i\partial )^{1+m}(\,\,i\partial ^{*})^{1+%
\widetilde{m}}\,\varphi
_{m,\widetilde{m}}^{1}(x,x^{*})=c\varphi
_{1-m,1-\widetilde{m}}^{2}(x,x^{*}),
\end{equation}
where $c$ is an unessential constant.
\\ \\
Starting from the symmetric solution (18), let us use this duality 
transformation in order to obtain an odderon solution. Since for odd values 
of the conformal spin $n$ the duality operator changes 
its sign under the transformation $x\longleftrightarrow 1-x$, the
duality transformations lead to relations between totally symmetric and
anti-symmetric wave functions $f(\mbf{\rho }_{1},\,%
\mbf{\rho }_{2},\,\mbf{\rho
}_{3};\,\mbf{%
\rho }_{0})$. In the particular case $\lambda =0$ the anti-symmetric wave
function $\varphi _{m,\widetilde{m}}^{0}(x,x^{*})$ is given
above in (22), and the symmetric wave function $\varphi
_{m,\widetilde{m}}^{odd}(%
x,x^{*})$ describing an odderon state can be obtained from the
solution of the equation
\begin{equation}
Q_{m,\widetilde{m}}\,\varphi _{m,\widetilde{m}}^{odd}(x,x^{*}%
)=c\varphi _{1-m,1-\widetilde{m}}^{0}(x,x^{*}).
\end{equation}
It is important to take into account that the solution 
$f_{m,\widetilde{m}}^{odd}(%
\mbf{\rho }_{1},\,\mbf{\rho
}_{2},\,\mbf{%
\rho }_{3};\,\mbf{\rho }_{0})$ includes the propagators of the 
external gluons. The amputated solution $F_{m,%
\widetilde{m}}^{odd}(\mbf{\rho }_{1},\,\mbf{\rho }%
_{2},\,\mbf{\rho }_{3};\,\mbf{\rho }_{0})$ with the
removed propagators can be written as follows
\[
F_{m,\widetilde{m}}^{odd}(\mbf{\rho
}_{1},\,\mbf{\rho }%
_{2},\,\mbf{\rho }_{3};\,\mbf{\rho }_{0})=\left|
\frac{%
1}{\rho _{12}\rho _{23}\rho _{31}}\right| ^{2}\left| A\right| ^{2}f_{m,%
\widetilde{m}}^{odd}(\mbf{\rho }_{1},\,\mbf{\rho }%
_{2},\,\mbf{\rho }_{3};\,\mbf{\rho }_{0})=
\]
\begin{equation}
\left| \frac{1}{\rho _{12}\rho _{23}\rho _{31}}\right| ^{2}\left(
\frac{\rho
_{23}}{\rho _{20}\rho _{30}}\right) ^{m}\left( \frac{\rho _{23}^{*}}{\rho
_{20}^{*}\rho _{30}^{*}}\right) ^{\widetilde{m}}\Phi _{m,\widetilde{m}%
}^{odd}(x,x^{*})\,,
\end{equation}
where $\Phi _{m,\widetilde{m}}^{odd}(x,x^{*})$ is obtained to be
\begin{equation}
\Phi _{m,\widetilde{m}}^{odd}(x,x^{*})=\left|
Q_{m,\widetilde{m}%
}\right| ^{2}\varphi _{m,\widetilde{m}}^{odd}(x,x^{*})=c\left|
x(1-x)\right| ^{2}(i\partial )^{2-m}(\,\,i\partial
^{*})^{2-\widetilde{m}%
}\varphi _{1-m,1-\widetilde{m}}^{0}(x,x^{*}).
\end{equation}
\\ \\ 
With the use of a Fourier transformation one can verify that
\begin{equation}
(i\partial )^{2-m}(\,\,i\partial ^{*})^{2-\widetilde{m}}\varphi _{1-m,1-%
\widetilde{m}}^{0}(x,x^{*})=a\left( \delta ^{2}(x)-\delta
^{2}(1-x)+\frac{x^{m}x^{*\widetilde{m}}}{\left| x\right| ^{6}}\delta
^{2}(%
\frac{1}{x})\right) ,
\end{equation}
where $a$ is a constant. Therefore we obtain
\[
F_{m,\widetilde{m}}^{odd}(\mbf{\rho
}_{1},\,\mbf{\rho }%
_{2},\,\mbf{\rho }_{3};\,\mbf{\rho }_{0})\sim
\]
\[
\left| \frac{\rho _{20}\rho _{30}}{\rho _{10}^{2}\rho _{32}^{3}}\right|
^{2}\left( \frac{\rho _{23}}{\rho _{20}\rho _{30}}\right) ^{m}\left(
\frac{%
\rho _{23}^{*}}{\rho _{20}^{*}\rho _{30}^{*}}\right)
^{\widetilde{m}}\left(
\delta ^{2}(x)-\delta ^{2}(1-x)+\frac{x^{m}x^{*\widetilde{m}}}{\left|
x\right| ^{6}}\delta ^{2}(\frac{1}{x})\right) \sim
\]
\begin{equation}
\frac{E_{m\widetilde{m}}(\rho _{20},\rho _{30})}{\left| \rho _{23}\right|
^{4}}\delta ^{2}(\rho _{12})+\frac{E_{m\widetilde{m}}(\rho _{10},\rho
_{20})%
}{\left| \rho _{12}\right| ^{4}}\delta ^{2}(\rho
_{31})+\frac{E_{m\widetilde{%
m}}(\rho _{30},\rho _{10})}{\left| \rho _{31}\right| ^{4}}\delta
^{2}(\rho
_{23}),
\end{equation}
where
\begin{equation}
E_{m\widetilde{m}}(\mbf{\rho }_{10},\mbf{\rho }%
_{20})=\left( \frac{\rho _{12}}{\rho _{10}\rho _{20}}\right) ^{m}\left(
\frac{\rho _{12}^{*}}{\rho _{10}^{*}\rho _{20}^{*}}\right)
^{\widetilde{m}}
\end{equation}
is the BFKL wave function. Using the fact that $E_{m\widetilde{m}}(%
\mbf{\rho }_{10},\mbf{\rho }_{20})$ is the 
eigenfunction of the Kasimir operators $M^{2}=\rho _{12}^{2}\partial
_{1}\partial
_{2}$ and $M^{*2}$ of the Moebius group, we can write the odderon 
solution (26) as follows:
\begin{equation}
F_{m,\widetilde{m}}^{odd}(\mbf{\rho
}_{1},\,\mbf{\rho }%
_{2},\,\mbf{\rho }_{3};\,\mbf{\rho }_{0})\sim
\sum_{i,k\neq l}\delta ^{2}(\mbf{\rho} _{li})\left| \mbf{\partial}_{i}\right|
^{2}\left| \mbf{\partial}_{k}\right|
^{2}E_{m\widetilde{m}}(\mbf{\rho }%
_{i0},\mbf{\rho }_{k0})\,,
\end{equation}
where the summation 
is performed over all gluon indices $i,\,k,\,l=1,2,3$
providing that $i,k\neq l$. After the transition to the momentum space we
obtain for this vertex function
\begin{equation}
F_{m,\widetilde{m}}^{odd}(\mbf{k}_{1},\,\mbf{k}_{2},\,
\mbf{k}_{3}) \sim \sum_{i,k\neq l}\left| \mbf{k}_{i}+\mbf{k}_{l}\right|
^{2}\left|
\mbf{k}_{k}\right|
^{2}E_{m\widetilde{m}}(\mbf{k}_{i}+\mbf{k}%
_{l},\mbf{k}_{k})\,,
\end{equation}
where
\begin{equation}
E_{m\widetilde{m}}(\mbf{k}_{1},\mbf{k}_{2})=\int
\frac{
d^{2}\rho _{1}d^{2}\rho _{2}}{(2\pi )^{2}}\exp \left( i\sum_{r=1}^{2}%
\mbf{k}_{r}\mbf{\rho }_{r}\right)
E_{m\widetilde{m}}(%
\mbf{\rho }_{1},\mbf{\rho }_{2})\,.
\end{equation}
Eq.(29) is the amputated counterpart of (9).\\ \\ 
%%%%%%%%%%%%%%%%%%%%%%%%%%%%%%%%%%%%%%%%%%%%%%%%%%%%%%%%%%%%%%%%%%%%%%%%
5. Finally let us analyze the coupling of the
new odderon state represented by its eigenfunction (9) to the
$\gamma^* \to \eta_c$ impact factor (3). Its knowledge
will permit us to study scattering processes with odderon exchange, using
the odderon Green function.
The key point to note is that only the full impact factor
$\Phi_{\gamma \to \eta_c}$ in (3) has a
''good'' infrared behaviour: it vanishes as any $\mbf{k}_i \to 0$.
Therefore, inside an integral any individual term $\varphi_{-}$ will have 
infrared singularities,
but they will cancel if we consider the full sum (3).
These singularities are also related to the nature of the conformal
invariant eigenfunction of the
BFKL pomeron (27): in the momentum representation they contain $\delta$-like 
pieces \cite{MT,BFLLRW}, corresponding to a constant behaviour in 
configuration space, as one of the
two coordinates $\mbf{\rho}_1$, $\mbf{\rho}_2$ is taken to $\infty$.
In a mathematical sense, therefore, the Pomeron eigenfunction is a 
distribution, and its meaning has to
be understood by integrating with some test function. Depending on the space 
of test functions, one can expect slightly different results. 
The same must also be true for the action of an 
operator on (27), e.g. the BFKL Hamiltonian:
the result will, again, be a distribution and has to be integrated with a test
function. All this is not a merely mathematical observation, but it has a 
natural physical meaning: the space of test functions in BFKL dynamics 
is defined by couplings (''impact factors'') to colorless objects. 
In (3) the function $\varphi_{-}$ 
alone does not have the normal ''good'' properties of a colourless object; only
the sum of all the terms in (3) defines a ''good'' function. \\ \\
Let us take a closer look at the scalar product of 
$\Phi_{\gamma \to \eta_c}$ with
$E_3^{(\nu,n)}$, taking into account the antisymmetry properties of the
two building objects, $\varphi_{-}$ and $E^{(\nu,n)}$. Using the 
momentum structure in (3) and (9) one finds
\begin{eqnarray}
\langle \Phi_{\gamma \to \eta_c} | E_3^{(\nu,n)} \rangle &=&
\int d\mu_3 \Phi_{\gamma \to \eta_c}(\{ \mbf{k}_i\})
E_3^{(\nu,n)}(\{ \mbf{k}_i\}) \nonumber \\
&=& -6 \int d^2 \mbf{k} \,
\Bigl[ \varphi_{-}(\mbf{k},\mbf{q}-\mbf{k}) - \varphi_{-}(\mbf{0},\mbf{q})\Bigr]
\left( \mbf{K}_L \otimes E^{(\nu,n)} \right) (\mbf{k},\mbf{q}-\mbf{k}) \, ,
\end{eqnarray}
where 
$d\mu_3=\prod_i d^2  \mbf{k}_i \, \delta^{(2)}(\mbf{q}-\sum_i\mbf{k}_i)$, and
\begin{eqnarray}
&&\left( \mbf{K}_L \otimes E^{(\nu,n)} \right) (\mbf{k},\mbf{q}-\mbf{k})=
\nonumber \\
&&\frac{N_c \alpha_s}{2 \pi^2}
\int d^2 \mbf{l} \Bigl[ \frac{\mbf{l}^2}{\mbf{k}^2(\mbf{l}-\mbf{k})^2}
E^{(\nu,n)}(\mbf{l},\mbf{q}-\mbf{l})- 
\frac{1}{2}\frac{\mbf{k}^2}{\mbf{l}^2(\mbf{k}-\mbf{l})^2}
E^{(\nu,n)}(\mbf{k},\mbf{q}-\mbf{k}) \Bigr] \, .
\end{eqnarray}
One sees easily that $\mbf{K}_L$ stands for ''half'' of the forward BFKL 
kernel; adding the corresponding expression for $\mbf{K}_R$ one obtains the
BFKL kernel, but without the local piece.
Due to the antisymmetry of $E^{(\nu,n)}$, the local term
in the BFKL kernel is giving zero contribution.
Therefore, in the first term of the integrand in (31), 
using the antisymmetry of $\varphi_{-}$, one could also include one half of 
this local BFKL-piece, such that $\mbf{K}_L$ really represents ''half'' of the
full BFKL kernel. Ignoring all potential divergences, one might naively expect
that (32) should, basically, lead to $\chi E^{(\nu,n)}$, and our scalar product
$\Phi \otimes E^{(\nu,n)}$ equals $\chi \varphi_{-}\otimes E^{(\nu,n)}$.
This expectation turns
out to be correct, but the argument is rather subtle. First, one notices    
that the first and the second integrand in (32) by themselves lead to
infrared divergent integrals, whereas the scalar product of
$\varphi_{-}$ with $E^{(\nu,n)}$ is convergent. So it is clear
that in the integration we are not allowed simply to use 
$\mbf{K}_{BFKL} \otimes E^{(\nu,n)} = \chi_{\nu,n} E^{(\nu,n)}$. The 
divergent pieces and, possibly also finite parts, would be lost. All these 
complications would not be visible if $\varphi_{-}$ would be a ''good'' 
function.\\ \\ 
In order to see the resolution to this puzzle, we calculate 
$\mbf{K}_L \otimes E^{(\nu,n)}$ explicitly. Rather than presenting details
of this calculation we only quote the main result. In the coordinate representation 
we find:
\beqn
(\mbf{K}_L \otimes E^{(\nu,n)})(\rho_1,\rho_2)=\frac{1}{2}\chi_{\nu,n}
E^{(\nu,n)}(\rho_1,\rho_2) + 
C \lim_{\rho_1 \to \infty}E^{(\nu,n)}(\rho_1,\rho_2) \, ,
\eeqn
where $C$ contains some infinite (after removing the infrared regularization) 
contributions and also finite pieces. In momentum space, this relation 
corresponds to the presence of some extra 
$\delta$ function-like pieces. 
It turns out that, as it should be, these extra terms give a
contribution which is exactly cancelled by  
the second integrand in (31) (the subtraction term). It is only after these
cancellations that finally we can write
\beqn 
\langle \Phi_{\gamma \to \eta_c} | E_3^{(\nu,n)} \rangle=
\sqrt{-3 \chi_{\nu,n}} \int d^2 \mbf{k} \,
\varphi_{-}(\mbf{k},\mbf{q}-\mbf{k}) E^{(\nu,n)} (\mbf{k},\mbf{q}-\mbf{k})\, .
\eeqn
Thus the matrix element of our odderon solution is similar to the 
corresponding matrix element of the pomeron solution: this is related to our
interpretation of the new odderon as a compound state of ''f'' and 
''d'' reggeized gluons. Note that the degeneracy between the ''f'' and the 
''d'' gluons is exact in the large $N_c$ limit, and the duality ~\cite{Ldual}
can be considered as a manifestation of this symmetry.   
Finally we just remark that in calculating the norm of $E_3^{(\nu,n)}$ the
$\delta$-like pieces do not play any role. In fact, in place of
$\Phi_{\gamma \to \eta_c}$ one has the amputated odderon function
given in (29) and, therefore, in place of $\varphi_{-}$
the amputated pomeron eigenfunction which has ''good'' properties.
\\ \\
%%%%%%%%%%%%%%%%%%%%%%%%%%%%%%%%%%%%%%%%%%%%%%%%%%%%%%%%%%%%%%%%%%%%%%%
6. We have presented a new set of eigenfunctions of the odderon equation
which is characterized by a spectrum with a maximum intercept at $1$.
It is remarkable that symmetry structure of this solution has been suggested 
by the impact factor $\Phi_{\gamma \to \eta_c}$, to which the odderon couples,
and by the Pomeron$\to$ two odderon vertex $W$ which came our from the 
study of the six gluon amplitude. At the same time one can use 
the duality symmetry of the 3 gluon Hamiltonian to
rederive this solution. This derivation shows the interconnection
with solutions of different symmetry properties.  
Finally we have shown how to calculate the scalar product of the
eigenfunction with the impact factor.
This opens the possibility to calculate numerically the contribution
of the new odderon states to some of those processes which have already
been studied to probe the QCD odderon. Work in this direction is in progess.
\\ \\ 
%%%%%%%%%%%%%%%%%%%%%%%%%%%%%%%%%%%%%%%%%%%%%%%%%%%%%%%%%%%%%%%%%%%%%%%
{\bf Acknowledgements} \\
J.B. and G.P.V. are grateful to M.A. Braun for very interesting discussions.
%%%%%%%%%%%%%%%%%%%%%%%%%%%%%%%%%%%%%%%%%%%%%%%%%%%%%%%%%%%%%%%%%%%%%%%
%References...
    
\end{document}